# Temperature distribution function of X–ray clusters of galaxies in a neutrino dominated universe

C. Balland[1,2] and A. Blanchard[1]

[1] Université Louis Pasteur, Observatoire Astronomique de Strasbourg, 11 rue de l'Université, F-67000 Strasbourg, France
[2] Université d'Aix-Marseille II, F-13009 Marseille Luminy, France

**Abstract.** The constraint that can be set on the Hot Dark Matter (HDM) model of structure formation from the temperature distribution function (TDF) of X–ray clusters of galaxies is examined in terms of the amplitude to be given to the primeval matter fluctuations for consistency with present observations.

The TDF of clusters is derived for different normalizations of the HDM power spectrum in an Einstein-de Sitter universe ($\Omega_0 = 1$) using the statistics of peaks of a random Gaussian field.

Concerning cluster formation only, the neutrino picture fails to reproduce the TDF for $H_0 = 50$ km s$^{-1}$ Mpc$^{-1}$ but appears to be marginally consistent with observations for $H_0 = 70$ km s$^{-1}$ Mpc$^{-1}$ provided the normalization is $\alpha \approx 0.42$, in terms of the ratio $\alpha = \sigma_0/\delta_s$ with $\sigma_0$ the rms variance of the fluctuations linearly evolved to the present epoch and $\delta_s$ the collapse threshold required in structure formation models.

However, using the constraint that the formation epoch of QSO's puts on the HDM model, it appears that this normalization is too small to be consistent with quasar formation. Allowing higher values for the Hubble constant can alleviate this problem but leads to an unacceptable short age of the universe. Moreover, the constraint put by the recent measurement of the level of fluctuations in the cosmic background radiation by the COBE DMR experiment (Smoot et al. 1992) tends to reject models with high values of $H_0$. Despite the fact that the observational constraints do not appear as stringent as they were claimed to be in some previous analysis, we still conclude that the neutrino model fails in any case to account for the whole set of available observations in a consistent way.

## 1. Introduction

Much work has already been done on the once popular Hot Dark Matter (HDM) model of structure formation (e.g: Cowsik & McClelland 1973; Bond et al. 1980; Peebles 1982; Bond & Szalay 1983; Kaiser 1983; Frenk et al. 1983; White et al. 1983, 1984; Melott 1985, 1987; Buchert 1991; Zeng & White 1991). Several attractive features naturally arise in this scenario:

1) first of all, while whether they are massive or not is still an open question, neutrinos do exist. Stringent limits on the electron neutrino mass have been obtained by different methods but the constraints put on the muon and tau neutrino masses do not exclude a mass of a few tens of eV. While the standard Big Bang nucleosynthesis models predict that the total amount of baryonic material in the universe should not exceed about one tenth of the closure density of the universe, the existence of a non baryonic dark matter constituent like one species of massive neutrinos might reconcile the abundance of baryonic material predicted by the nucleosynthesis with the exigence of a density parameter $\Omega_0$ close to 1 as expected in inflationary scenarios.

2) the short-wavelength cut-off in the HDM power spectrum provides a natural explanation to the filamentary aspect of structure on large scale (de Lapparent et al. 1986) and to the large coherence observed in the real galaxy distribution.

However, the model seems not to resist to a closer investigation. As pointed out by Peebles (1982), then emphasized by the numerical experiments of White et al. (1983, 1984), the neutrino picture suffers from a *timing problem* that comes to light when comparing the large-scale structure properties at the present epoch with the small-scale constraint at high redshift. This problem can be conveniently expressed in terms of the normalization of the model through $\sigma_0$, the rms variance of the matter fluctuations linearly evolved until now. Indeed, each observational constraint can be related to a value required for $\sigma_0$. In general, a single observational constraint will be satisfied for a certain range of values of $\sigma_0$. A robust test of the model is obtained by examining simultaneously several constraints, i.e. the consistency of the values of $\sigma_0$ required by the various observations. In their pioneering work, White et al. (1984) choose the onset of galaxy formation as the time when one percent of the mass in their simulations has undergone collapse. This happens, in their simulations, after an expansion factor of 3, when the linear rms fluctuation is $\sigma = 0.66$. In order to take into account the existence of high redshift quasars ($z \geq 2$), they normalize their simulations to $\sigma_0 = 0.66(1 + z_c) \approx 2.2$ and $\sigma_0 \approx 4.4$ so that the first collapse of objects occurs at $z_c = 2.5$ and $z_c = 6$ respectively. As a result of using such high normalizations, the clusters produced later in the simulations, i.e. at $z = 0$, have a much higher mass (and temperature) than the one observed in the real galaxy distribution. From the argument that HDM cannot reproduce consistently both the small-scale, i.e QSO's, and the large-scale, i.e. clusters, properties of structure, the model was early ruled out (White et al. 1984; Kaiser 1983). However, it should be emphasized that the choice of the onset of QSO's formation is quite uncertain as they represent a negligible fraction of the total collapsed material and are below the resolution limit of White et al. numerical experiments. In retrospect, their criterion may seem too drastic and their conclusions have thus been questioned (Melott 1985; Melott 1987; Centrella et al. 1988); accordingly, the model has been revisited from time to time (Anninos et al. 1991; Zeng & White 1991; Cen & Ostriker 1992; Tuluie et al. 1994).

Recently, Blanchard et al. (1993) have revisited the constraints set from small-scale (QSO's) observations in the neutrino picture. Their semi-analytical approach leads to the conclusion that, considering the large uncertainty that exists on the density threshold $\delta_s$ involved in structure formation models, the existence of high redshift quasars can be taken into account if the normalization of the spectrum, $\sigma_0$, ranges from 1 to about 3. The normalizations used in the pioneering work of White et al. (1984) clearly do not cover all this range.

At this point, it seems interesting to us to revise the constraint that can be set on the HDM model from galaxy clusters. In this paper, we propose an analytic approach using the cluster temperature distribution function (TDF), i.e. the abundance of galaxy clusters as a function of their X–ray temperature, to constrain the normalization of the HDM spectrum. The constraints that can be set on galaxy formation theories from the TDF of clusters have been discussed in various models (Henry & Arnaud 1991; Blanchard & Silk 1991; Bartlett & Silk 1993; Oukbir & Blanchard 1992), while some authors have rather used the constraint imposed by the velocity distribution function (Evrard 1989, Lilje 1990). However, this latter method suffers from uncertainties as cluster velocity dispersion might be overestimated (Frenk et al. 1990) because of projection effects along the line of sight. Moreover, the observed galaxy velocity dispersion, $\sigma_{gal}$, might not trace the matter velocity dispersion: there might exist a velocity bias as claimed by Carlberg & Couchman (1989) according to their simulation results. On the contrary, X–ray data appear to be more reliable (Edge et al. 1990). The X–ray luminosity function has been also used to constrain galaxy formation models (Kaiser 1986, Evrard & Davis 1988). However, the luminosity is entirely dominated by the core properties of the X–ray gas that has been shown to reside in clusters and it is unlikely that its scaling properties can be modeled confidently (Blanchard et al. 1992b). On the contrary, the gas temperature is a quite direct measure of the potential depth and the X–ray temperature distribution function appears as a clean test.

In our approach, the main hypothesis is that the cluster gas is close to isothermal hydrostatic equilibrium at least out to the virial radius. Observations tend to confirm isothermality in the central part (apart from clusters in which evidence of cooling flows has been found). Hydrodynamical simulations from Evrard & Davis (1988) tend also to favour this hypothesis and we adopt it hereafter.

Our derivation of the cluster TDF is based on the statistics of peaks of the initial matter density field (see Bardeen et al. 1986, BBKS throughout this paper) and on the spherical collapse model (Lemaître 1933; Gunn & Gott 1972; Peebles 1980). This method appears well designed for an HDM scenario since peaks of the field expected to lead to structure formation are unambiguously defined (because of the absence of structure on small scales), whilst the same method for the Cold Dark Matter model, as for any hierarchical scenario, faces the so-called 'cloud-in-cloud' problem when the field is smoothed from small scales to larger scales (BBKS, Peacock & Heavens 1985). Using the spherical infall model is however more questionable as in a 'pancake' process the collapse of matter is expected to proceed along one privileged axis leading to the formation of highly anisotropic structures. The collapse of peaks might occur for a density threshold $\delta_s$ smaller than the canonical value 1.68 derived in the top-hat model. However, high-$\nu$ peaks are expected to be spherical (BBKS). Moreover, it has been shown that high-$\nu$ density fluctuations rigorously follow the spherical collapse for any cosmological model (Bernardeau 1994), so that this value, $\delta_s = 1.68$, seems to be fair. Nevertheless, a possible way to get rid of the systematic uncertainty on $\sigma_0$ induced by the uncertainty on the threshold $\delta_s$ is to express the derived normalizations in terms of the parameter $\alpha$ (the inverse of the lowest $\nu$ of structures that form):

$$\alpha = \sigma_0/\delta_s$$

rather than merely $\sigma_0$. Blanchard et al. (1993) pointed out that the quantity $\alpha$ is directly constrained by the observations independently of the uncertainty on $\delta_s$. This will allow us to compare in a consistent way the normalization we obtain on cluster scale to the one derived by Blanchard et al. for QSO's.

## 2. Cosmological framework and fluctuation spectrum

We consider the growth of large scale inhomogeneities of the initial matter density field in an Einstein-de Sitter ($\Omega_0 = 1$) universe according to gravitational instability theory (in the following, the subscript 0 for the density parameter and the Hubble constant refers to present values). The statistical properties of the field, assumed to be random and Gaussian, are described by $P(\mathbf{k}, t)$, the power spectrum of the fluctuations, i.e., the correlation function of the Fourier transform of the density contrast $\delta$ defined as the local excess of matter over the mean density of the universe:

$$P(\mathbf{k}, t) = \left\langle |\delta(\mathbf{k}, t)|^2 \right\rangle, \qquad \delta(\mathbf{r}) = \frac{\rho(\mathbf{r}) - \langle \rho \rangle}{\langle \rho \rangle}. \qquad (1)$$

In the HDM model, the assumption is made that the universe consists mainly of massive neutrinos and in an $\Omega_0 = 1$ cosmology, according to primordial nucleosynthesis, neutrinos must then account for 95% of the total matter if $H_0 = 50$ km s$^{-1}$ Mpc$^{-1}$. Such an assumption constrains the power spectrum except for its normalization (the amplitude of the primordial fluctuations) which is the only main free parameter of the theory. It is usual to choose the Harrison-Zel'dovitch scale invariant spectrum as the initial spectrum: $P_i(\mathbf{k}, t) \propto k^n$ with $n = 1$. The power spectrum at recombination is obtained by multiplying $P_i$ by a transfer function that accounts for the history of the fluctuations (free streaming damping in the neutrino hypothesis). We choose the transfer function derived from Bond & Szalay (1983) and given by BBKS. The inferred spectrum is characterized by a cut-off $\lambda_c$ which depends only on the density parameter of the neutrino constituent $\Omega_\nu$ (for one species) and the Hubble constant $H_0 = 100h$ km s$^{-1}$ Mpc$^{-1}$:

$$\lambda_c \approx 13 \, (h^2 \Omega_\nu)^{-1} \text{ Mpc}.$$

We approximate $\Omega_\nu \approx 1$. The spectrum presents a lack of power at small scales, which prevents small structures such as galaxies from forming first. Clusters then appear as the first structures to form in this model.

## 3. Temperature distribution function of galaxy clusters

In the gravitational instability picture the maxima of the matter density field are likely to be the privileged sites of collapse of non-linear objects such as clusters of galaxies (BBKS 1986). In hierarchical models, the counting of peaks is not reliable because of the 'cloud-in-cloud' problem and a Press & Schechter-like approach turns out to be more adapted (Blanchard et al. 1992a). In the HDM scenario, contrary to hierarchical scenarios, there is one definite epoch at which the first structures form. This epoch corresponds to the collapse of the highest peaks of the field. It is only much later, when $\sigma_0 \gg 1$, that fluctuations on large scales will collapse, embedding the first generation objects. In this scenario, the count of non-linear objects from the peaks of the density field is thus a reliable approach at least during the collapse of the first objects (see below for a more quantitative discussion of the 'cloud-in-cloud' in the HDM model). During the collapse of the first peaks, the gas is shock-heated, reaches the virial temperature and is gravitationally confined through an hydrostatic equilibrium. The final temperature depends on the total collapsed mass. In order to identify clusters hotter than $T$, i.e. more massive than some value $M$, we first smooth the density field by a top-hat window function of radius $R$ (with $M = \bar{\rho} \frac{4}{3} \pi R^3$) and we assume that clusters form at location where the smoothed field is above the threshold $\delta_s$. The choice of an adequate filtering window for smoothing the field is one difficulty in this approach. The spherical (top-hat) window is probably the most relevant if the collapse of structure is expected to be spherical. However, the sharp features of the window in real space induces oscillations in Fourier space, which adds power to the spectrum out to several cut-off scales. The Gaussian window avoids such a difficulty but the relation between the mass of a collapsed object and the mass of the corresponding peak in the field is not obvious in this case (Peacock & Heavens 1985). However this problem is of minor importance for HDM since the low-pass filtering procedure induces high-frequency spikes on scales where there is no power. We therefore expect that Gaussian and top-hat filtering lead to comparable results in this case.

The temperature distribution function $\mathcal{N}(T)$ of clusters at some epoch $z_v$ can then be directly related to the statistics of peaks of the smoothed field:

$$\mathcal{N}(>T) = \int_{\nu_s(T)}^{\infty} \mathcal{N}_{pk}(\nu, R_\star) d\nu = \int_T^\infty \mathcal{N}(\bar{T}) d\bar{T} \qquad (2)$$

In equation (2) we use the formalism developed by BBKS, so that $\mathcal{N}_{pk}$, $\nu$ and $R_\star$ are defined by:

$$\mathcal{N}_{pk}(\nu, R_\star) = \frac{1}{(2\pi)^2 R_\star^3} e^{-\nu^2/2} G(\gamma, \gamma\nu) \qquad (3.1)$$

$$\nu = \frac{\delta}{\sigma_0}(1 + z_v), \quad R_\star \equiv \sqrt{3}\frac{\sigma_1}{\sigma_2} \qquad (3.2)$$

where $G(\gamma, \gamma\nu)$ in equation (3.1) is the function given by equation (A.19) of BBKS and the parameter $\gamma$ is defined by:

$$\gamma \equiv \frac{\sigma_1^2}{\sigma_2 \sigma_0}. \qquad (3.3)$$

$\gamma$ and $R_\star$ are called the spectral elements of the field and they involve the spectral moments $\sigma_i$ defined below:

$$\sigma_i^2(t) \equiv \int \frac{k^2 dk}{2\pi^2} P(k, t) k^{2i}. \qquad (4)$$

$\sigma_0(R_f)$ is the rms fluctuation of the field filtered at the comoving scale $R_f$. $\sigma_1(R_f)$ and $\sigma_2(R_f)$ are related to the first and second derivative of the field. $\mathcal{N}_{pk}(\nu)$ is the mean differential number of peaks per comoving volume unit (cf BBKS). $R_\star$ can be related to the mean distance between peaks in the following way: the cumulative number density of peaks of arbitrary height is:

$$n_{pk}(-\infty) = \int_{-\infty}^{\infty} \mathcal{N}_{pk}(\nu) d\nu \approx 0.016 R_\star^{-3}. \text{ (BBKS : 4.11}b)$$

The mean distance between peaks of arbitrary height is then:

$$<d_{pk}> = (0.016 R_\star^{-3})^{-\frac{1}{3}} \sim 4 \times R_\star.$$

The approach considered in equation (2) has been known to be plagued, in a hierarchical scenario of structure formation, by the so-called cloud-in-cloud problem (BBKS 1986, Peacock & Heavens 1985). This term recovers the fact that two (or more) separate peaks in the field smoothed at some scale $R_1$ might be part of a single peak, i.e. structure, at some other scale $R_2 > R_1$. Such an approach is therefore likely to overestimate the real number of structures in a hierarchical scenario since several non-linear peaks fated to end up into a single structure might be counted as individual objects. This is probably the major issue that makes the 'peak approach' a crude way of estimating the temperature function in this kind of scenario.

In the case of a top-down type model as the HDM model envisaged here, because of the lack of power on small scales, the mean distance between peaks is large enough at all the filtering scales of interest, i.e. on cluster scales, to avoid the possibility that separate peaks on one scale are absorbed in a unique structure on a larger scale. The mean distance between peaks of arbitrary height is $4 \times 13.69 \sim 55$ Mpc ($h = 0.5$) for HDM and the mean distance between peaks on cluster scale ($8h^{-1}$ Mpc) is $\sim 93$ Mpc. We thus expect that peaks in the HDM scenario are unambiguously defined and that equation (2) gives a good approximation to the number of non-linear objects.

Deriving this equation leads to:

$$\mathcal{N}(T) = \mathcal{N}_{pk}(\nu_s, R_\star)\frac{d\nu_s}{dT} - [\int_{\nu_s}^{\infty} \frac{\partial \mathcal{N}_{pk}(\nu, R_\star)}{\partial R_\star} d\nu]\frac{dR_\star}{dT}. \quad (5)$$

The first term in the second member of equation (5) counts the number of peaks with height $\nu_s$ within $d\nu_s$ while the second term counts the variation in the number of peaks above the threshold $\delta_s$ when the smoothing scale changes so that $R_\star$ increases of $dR_\star$. As it will be clear on figure 1 below, the second term can be neglected on cluster scales so that equation (5) can be approximated by:

$$\mathcal{N}(T) \sim \mathcal{N}_{pk}(\nu_s, R_\star)\frac{d\nu_s}{dT} \quad (6)$$

which can be rewritten, according to (3.2), as:

$$\mathcal{N}(T) \sim \mathcal{N}_{pk}(\nu_s, R_\star) \delta_s (1+z_v) \frac{-1}{\sigma_0^2} \frac{d\sigma_0}{dT}. \quad (7)$$

Note that this equation depends on $\alpha = \sigma_0/\delta_s$ and not on $\sigma_0$ or $\delta_s$ separately so that it is not necessary to specify the value of $\delta_s$ at this level. In equation (7), $\sigma_0$ is related to the temperature through the smoothing scale $R_{TH}$ and:

$$\frac{d\sigma_0}{dT} = \frac{d\sigma_0}{dR_{TH}} \times \frac{dR_{TH}}{dT}. \quad (8)$$

The scale $R_{TH}$ is related to the mass of the peak by:

$$M \sim \frac{4\pi}{3} R_{TH}^3 \bar{\rho} \quad (9)$$

where $\bar{\rho}$ is the mean density of the universe and a relation between the mass and the temperature of a newly formed object (i.e. galaxy cluster) is needed to evaluate $dR_{TH}/dT$. We derive this relation from the assumption that the intracluster medium, the baryonic gas that has been shown to reside in clusters, is in an hydrostatic equilibrium state at constant temperature. Such an assumption is discussed below. In this case, the equation of hydrostatic equilibrium for the gas is merely:

$$k_B T = \frac{-\mu m_p}{\gamma_\rho + \gamma_T} \frac{GM}{R} \quad (10)$$

$$\gamma_\rho = \frac{d\ln\rho_g}{d\ln R}, \qquad \gamma_T = \frac{d\ln T}{d\ln R} = 0$$

where $\mu$ is the mean molecular weight of a fully ionized plasma ($\mu \approx 0.63$), $k_B$ is the Boltzmann constant, $m_p$ is the proton mass, $\rho_g$ is the intracluster gas density and $M$ is the mass enclosed in a spherical shell of radius $R$. Equation (10) leads to: $T \propto M_{15}^{2/3}(1+z_v)h^{2/3}$ keV with $z_v$ the epoch of formation and $M_{15}$ the cluster spherical mass, related to the smoothing scale as mentioned above (equation (9)), in unit of $10^{15}$ solar mass. We normalized this relation to Evrard's hydrodynamic simulations result (Evrard 1990) and we obtained:

$$T \approx 6.4 M_{15}^{2/3}(1+z_v)h^{2/3} \text{ keV}. \quad (11)$$

Evrard's numerical simulations have been performed for a CDM initial power spectrum and using them for the neutrino picture is questionable. However, the derivation of (11) involves mainly the equation of hydrostatic equilibrium and the expression of the mass enclosed in a sphere of $\Delta \approx 200$, the density contrast of a virializing spherical cluster according to the spherical collapse model. As these two relations are independent on the characteristics of the underlying fluctuation spectrum, and as cluster collapse appears to be insensitive to the amount of small-scale power present in the spectrum (see Evrard & Crone (1992) simulations), we expect (11) to remain valid in the neutrino scenario. Lilje (1990) made the same assumption. Let us mention again that equation (11) assumes that the gas in galaxy clusters can be regarded as isothermal. This assumption is favoured by observations (Hughes et al. 1988; Hughes 1992) but is uncertain far from cluster cores.

The only remaining quantity to be determined in equation (7) is the ratio of the overall amplitude of the fluctuations ($\sigma_0$ at 0 Mpc), to the threshold $\delta_s$, i.e. the parameter $\alpha$. Note that $\sigma_0$, which is nothing but an integral over the fluctuation spectrum according to definition (4), diverges in 0 for CDM but converges for HDM. We propose to determine the quantity $\alpha$ by scaling relation (7) on the observed temperature function of galaxy clusters given by Edge et al. (1990). This is one of the main purpose of this paper.

## 4. Results and discussion

We present on figure 1 the temperature distribution function of galaxy clusters as obtained from equation (5) (solid line) and from equation (6) (dashed line). The curves are plotted for different redshifts of cluster formation. The normalization of the spectrum is arbitrary and has been taken to be $\alpha \approx 0.36$. The graphs are given for $h = 0.5$ which is required in order to obtain an acceptable age of the universe. Because of a normalization in the linear regime, strong evolution effects between redshift 1 and now can be noticed. The main feature of the temperature distribution function is a plateau for temperatures below 1 keV, which corresponds to galaxies or small groups. This feature clearly results from the lack of power at small scales in the neutrino picture spectrum. In Figure 2, we present the temperature distribution function for different normalizations of the spectrum at $z_v = 0$ obtained from equation (6), compared to observations. The filled squares correspond to Exosat and Einstein observatory data on a sample of 55 X–ray clusters (Edge et al. 1990). The filled square at 13.6 keV is from M. Arnaud et al. (1992). Error bars are given at 90% confidence level. In order to fit correctly the cluster abundances with temperature in the range 5 to 15 keV, it is necessary to have the amplitude $\alpha$ in the range 0.33 to 0.39 (these bounds do *not* depend on the actual value of the threshold of collapse $\delta_s$). Higher amplitudes would copiously overproduce high temperature clusters. On smaller temperatures ($T < 5$ keV), the curve conflicts with data (points P1 and P2 on figure 2a). The observed number density of clusters at $T \approx 2.2$ keV (point P1) is $\approx 2.7 \ 10^{-6}$ Mpc$^{-3}$ while the total number density of clusters we derive at this temperature from the peak approach is $\approx 1.5 \ 10^{-7}$ Mpc$^{-3}$. In P2 ($T \approx 3.5$ keV), the observed number density is $\approx 1.4 \ 10^{-6}$ Mpc$^{-3}$ while we find $\approx 1.4 \ 10^{-7}$ Mpc$^{-3}$. For HDM, the total number of peaks (i.e. the number of clusters as in this scenario one peak leads to the formation of one cluster) with height $> \nu = \alpha^{-1} \approx (0.35)^{-1} \approx 2.85$ is (see figure 3 from BBKS with $\gamma \approx 0.74$ and $R_\star \approx 13.69$ Mpc): $\approx 6.2 \ 10^{-7}$ Mpc$^{-3}$. The total observed number density of clusters for P1 and P2 is $\approx 4.1 \ 10^{-6}$ Mpc$^{-3}$, about one order of magnitude higher than the total number density of clusters expected in the HDM scenario so that the HDM model can not reproduce the observed TDF for $h = 0.5$. How-ever, temperatures lower than 5 keV correspond to groups or small clusters scales which might form by fragmentation of pancakes in the HDM model. Structures formed by fragmentation of peaks have not been taken into account in our analysis and fragmentation processes may then explain the discrepancy between the curve and the data. We can test this possibility by the following procedure. Fragmentation may increase the number of clusters and as the 'new' clusters will be in the neighborhood of the 'parent peak', this will translate into a high correlation between clusters. From the observations, we can evaluate the mean number of clusters $N_v$ associated with a collapsed structure, i.e. the mean number of neighbours of a given cluster:

$$N_v = n_c \int_0^R \xi_c(r) 4\pi r^2 dr \qquad (12)$$

where $n_c$ is the comoving number density of clusters, $\xi_c(r)$ the cluster-cluster two-point correlation function, and $R$ the comoving radius of the associated peak in the field. The cluster-cluster correlation function reads:

$$\xi_c(r) = \left(\frac{r}{r_0}\right)^{-\gamma} \qquad (13)$$

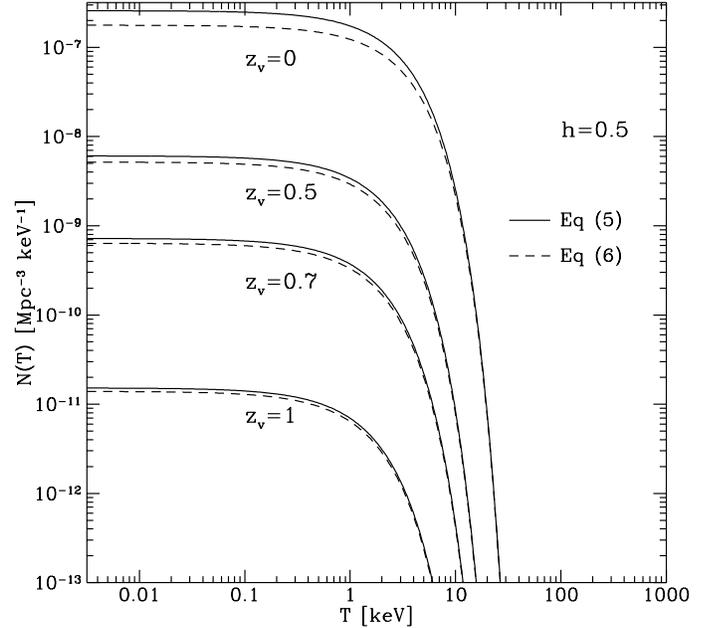

**Fig. 1.** Temperature distribution function from the peak approach for different redshifts. The spectrum is arbitrarily normalized to $\alpha \approx 0.36$. The solid line is obtained from equation (5) and the dashed line is from equation (6). The second term in equation (5) turns out to be negligible. As the spectrum is normalized in the linear regime, strong evolution effects are noticed between redshift 1 and the present epoch. The main striking feature of the TDF is a plateau for temperatures below 1 keV.

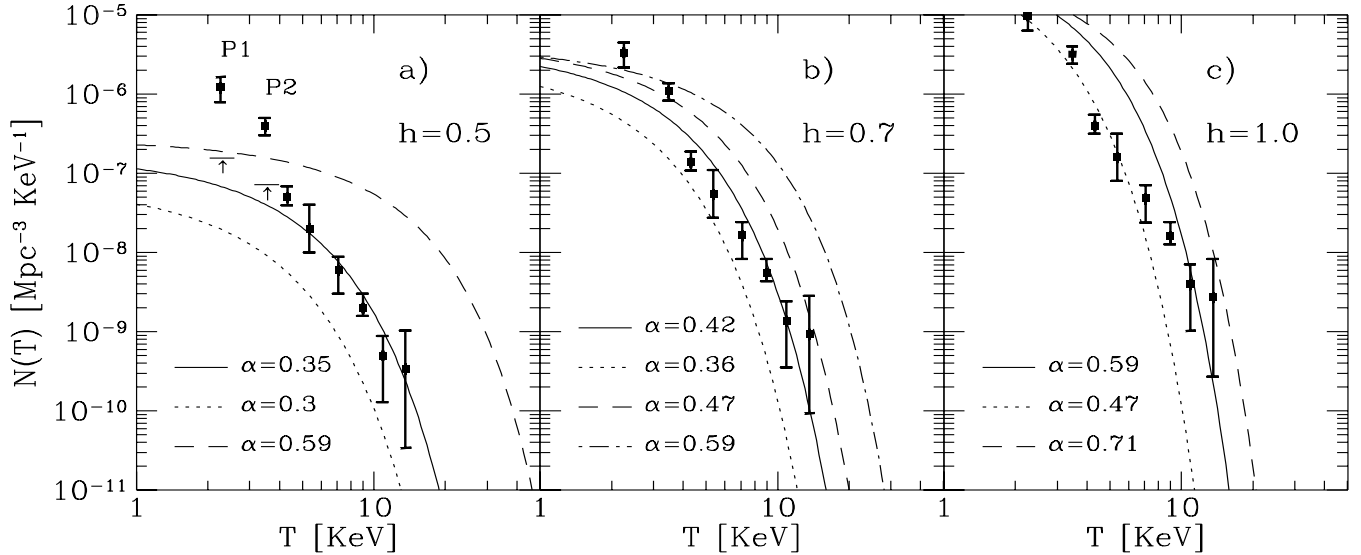

**Fig. 2. a-c** Temperature distribution function from the peak approach for different normalizations of the HDM spectrum compared to Edge's (1990) X–ray data: **a** $h = 0.5$, **b** $h = 0.7$, **c** $h = 1$. The cluster at 13.6 keV is from M. Arnaud et al. (1992). Error bars are given at 90% confidence level. The arrows on **2. a** denote the upper limits to which the solid curve can be pushed up in the presence of fragmentation.

with $r_0 \approx 25 h^{-1}$ Mpc and $\gamma \approx 1.8$ (Hauser & Peebles 1973). Using (13), equation (12) leads to:

$$N_v = 4\pi \frac{n_c}{(3-\gamma)r_0^{-\gamma}} R_\star^{3-\gamma} \qquad (14)$$

where the radius $R$ has been chosen to be roughly $R_\star$, which tends to overestimate $N_v$. For point P1 (figure 2a), $n_c$ is estimated to be $2.2 \times 1.2\ 10^{-6} \approx 2.7\ 10^{-6}$ Mpc$^{-3}$ at $T \approx 2.2$ keV and (14) leads to $N_v \approx 1.1$ for $h = 0.5$. For P2, we estimate $n_c \approx 1.4\ 10^{-6}$ Mpc$^{-3}$ and we find $N_v \approx 0.67$. In the presence of fragmentation, the actual number of peaks is:

$$\mathcal{N}_{pk} \times (1 + N_v)\ \text{Mpc}^{-3}. \qquad (15)$$

On figure 2a we put the upper limits to which the solid curve can be pushed up due to fragmentation processes. Fragmentation turns out not to be efficient enough to explain the discrepancy between the model and the observations. Note that the low-end slope of our model TDF should be greatly in error to reconcile the curve and the data.

We therefore conclude that the HDM model with $h = 0.5$ fails to reproduce the TDF of clusters. On figure 2b and 2c, the same graphs as figure 2a are presented for $h = 0.7$ and $h = 1$ respectively. While the normalization appears to be only slightly sensitive to the Hubble parameter, the fit is improved in the case of $h = 0.7$. For $h = 1$, the fit is no longer good and the age of the universe is unacceptably small. We find that for $\alpha \approx 0.42$ HDM with $h = 0.7$ reproduces roughly the observed cluster temperature function.

Figure 2 shows also that the TDF of clusters is very sensitive to the initial amplitude of matter fluctuations: increasing the normalization by a factor 2 results in about a factor 4 in temperature. In other words, the comparison between the TDF and observations provides a powerful constraint on $\alpha$. We draw on figure 2a and 2b the TDF for $\alpha = 0.59$. For $\delta_s = 1.68$ this corresponds to a normalization $\sigma_0 \approx 1$. It appears clearly that this normalization is inconsistent with the data (for $\delta_s < 1.68$, the discrepancy is reinforced). *A fortiori*, $\sigma_0$ values used by White et al. (1984) (cf introduction) are, as expected, much too large to account for cluster formation consistently with observations. Note that, as a check of our calculation, the number density of neutrino clusters we obtain at 40 keV with their normalizations is in good agreement with their simulation results.

The Press & Schechter (1974) formalism is not expected to provide a good estimation of the mass function in the HDM model. Specifically, Press & Schechter assumed that the total mass in non-linear objects with mass $M$ is related to the probability of finding a sphere that satisfies the non-linear criterion:

$$P(>\nu_s) = \int_{\nu_s}^{\infty} \frac{1}{\sqrt{2\pi}} e^{-\frac{\nu^2}{2}} d\nu, \qquad (16)$$

the field having been smoothed on the scale $R_0 = (3M/4\pi)^{\frac{1}{3}}$. This measures only the volume corresponding to the centers of such non-linear spheres while the full volume of spheres of radius $R > R_0$ should be taken into account; therefore (16) is an underestimate. The factor 2 introduced by Press & Schechter (1974) to correct this effect is purely phenomenological. However, as the Press & Schechter mass function is known to fit rather well the results of numerical simulations, we have checked to what normalization it leads. Using equation (11) and the Press & Schechter prescription for the mass function, we get:

$$\mathcal{N}(T) = \sqrt{\frac{2}{\pi}} \frac{\bar\rho}{M} \frac{\delta_s(1+z_v)}{\sigma_0^2} e^{-(\nu^2/2)} \left|\frac{d\sigma_0}{dT}\right|. \qquad (17)$$

This approach leads quantitatively to the same results as those on figure 2. We find, in order to fit Edge's data: $\alpha \approx 0.47$ ($h = 0.7$). This result is slightly higher than the normalization derived previously. This may reflect the fact that the Press & Schechter approach is likely to underestimate the true number of non-linear objects on the high mass end. In both cases however, the normalization obtained shows that according to HDM, most of the matter in the universe is still in the linear regime by now.

Until now, we have shown that the neutrino picture is able to reproduce consistently with observations the galaxy cluster distribution in an universe with $\Omega_0 = 1$ and $h = 0.7$ provided the normalization is adequately chosen. The validity of the model can be checked, as mentioned in the introduction of this paper, by comparing this normalization to the one obtained from QSO's by Blanchard et al. (1993). Recall that comparing directly the amplitude of $\alpha$ for clusters and QSO's cancels the uncertainty due to the unknown threshold $\delta_s$. It appears from the constraints put on quasar formation that $\alpha_{\rm QSO's}$ has to be greater than 1.7 ($h = 0.5$). In this case, the ratio of the normalizations in terms of $\alpha$ is:

$$\frac{\alpha_{\rm QSO's}}{\alpha_{\rm Clusters}} \approx 4.8. \qquad (18)$$

For higher values of the Hubble constant, this discrepancy can be reduced. For $h = 0.7$ and $h = 1$, $\alpha_{\rm QSO's} \approx 1.4$ and 1 respectively so that:

$$\frac{\alpha_{\rm QSO's}}{\alpha_{\rm Clusters}} \approx 3.3, \qquad \frac{\alpha_{\rm QSO's}}{\alpha_{\rm Clusters}} \approx 1.7. \qquad (19)$$

For $h = 1$, the discrepancy is much smaller but the two values are still inconsistent. However if we use the most conservative assumption on $\delta_s$ for QSO's, i.e. $\delta_s \approx 1$ while clusters follow the spherical infall model ($\delta_s = 1.68$), the normalizations in terms of $\sigma_0$ become consistent for $h = 1$.

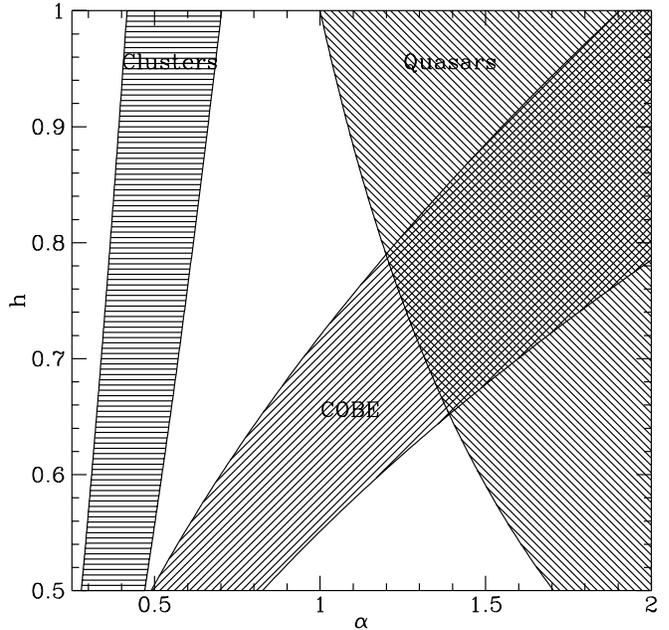

**Fig. 3.** The different constraints put on the HDM model plotted in the $\alpha - h$ space. The cluster constraint is derived from the present work while the QSO's constraint is obtained from Blanchard et al. (1993). The COBE constraint is evaluated from Górski et al. (1994) and Holtzman (1989). The hashed areas represent the allowed regions in the $\alpha - h$ space according to the corresponding constraint.

But the age of the universe in this case would be only approximately 6.5 Gyrs.

We summarize our results on figure 3 where the different constraints on the HDM model have been plotted in the $\alpha$-$h$ space. The cluster constraint has been derived from the work presented in this paper while the quasar constraint has been obtained from Blanchard et al. (1993). The COBE constraint is evaluated from Holtzman (1989) using the rms quadrupole coefficient value ($Q_{rms}$[1] $\approx 20 \pm 5$ $\mu$K($2\sigma$)) obtained by Górski et al. (1994) for a spectral index $n = 1$. For the cluster constraint, we allow the virial temperature to differ from the observed value by a factor of two so that the allowed range for the normalization $\alpha$ is substantially increased with respect to the range obtained previously. Even with this wide range of allowed values for $\alpha$, figure 3 shows that the HDM model in an Einstein-de Sitter universe can not account simultaneously for the different observational constraints.

---

[1] $Q_{rms} = \sqrt{\frac{5}{4\pi}} a_2$ where $a_2$ is the quadrupole coefficient in the spherical harmonic expansion of the sky temperature $\frac{\delta T}{T}$.

8	C. Balland & A. Blanchard: Temperature function of X–ray clusters in a neutrino dominated universe## 5. Conclusion

Concerning cluster formation only, we draw the following conclusions on the neutrino model:
1) the HDM model in an Einstein-de Sitter ($\Omega_0 = 1$) universe with $h = 0.5$ fails to reproduce the TDF.
2) HDM in an $\Omega_0 = 1$ universe with $h = 0.7$ is marginally consistent with the available observations provided that the normalization is adequately chosen: $\alpha \approx 0.42$ ($\sigma_0 \approx 0.7$ for $\delta_s = 1.68$). (Note that the age of the universe in this case is also marginally acceptable).
3) HDM in an Einstein-de Sitter universe with $h = 1$ is excluded by the temperature function constraint. (The age of the universe is also unrealistically low).

Concerning small-scale constraint, Blanchard et al. (1993) concluded that QSO's can be accommodated with a lower normalization than those used by White et al. (1984). However, we find that the *timing problem* still remains as both large scale and small scale constraints can not be satisfied simultaneously for a single amplitude of the primordial fluctuations. Higher Hubble constants reduce the discrepancy but lead to a very short age of the universe. Moreover, the level of CBR fluctuations as measured recently by COBE tends to exclude a model with a high Hubble constant. We conclude that in any case the neutrino picture has to be ruled out.

*Acknowledgements.* We thank the unknown referee for constructive comments that helped us to improve the manuscript of this paper. We also thank Jim Bartlett for helpful discussions. C.B. acknowledges finantial support from the French Ministère de la Recherche et de l'Enseignement Supérieur.## References

Anninos P., Matzner R.A., Tuluie R., Centrella J.M. 1991, ApJ 382, 71
Arnaud M., Hugues J.P., Forman W., Jones C., Lachieze-Rey M., Yamashita K., Hatsukade I. 1992, ApJ 390, 345
Bardeen J.M., Bond J.R., Kaiser N., Szalay A.S. 1986, ApJ 304, 15
Bernardeau F. 1994, ApJ 427, 51
Bartlett J.G., Silk J. 1993, ApJ 407, L45
Blanchard A., Silk J. 1991, Proc. 26th Rencontres de Moriond in Astrophysics (Gif-sur-Yvette: Editions Frontières) p. 93
Blanchard A., Valls-Gabaud D., Mamon G.A. 1992a, A&A 264, 365
Blanchard A., Wachter K., Evrard A.E., Silk J. 1992b, ApJ 391, 1
Blanchard A., Buchert T., Klaffl R. 1993, A&A 267, 1
Bond J.R., Efstathiou G., Silk J. 1980, Phys. Rev. Lett. 45, 1980
Bond J.R., Szalay A.S. 1983, ApJ 274, 443
Buchert T. 1991, In: *Physical Cosmology*, eds: Blanchard A. et al. , Frontières Paris, pp.475-483
Carlberg R.G., Couchman H.M.P. 1989, ApJ 340, 47
Cen R., Ostriker J.P. 1992, ApJ 399, 331
Centrella J.M., Gallagher III J.S., Melott A.L., Bushouse H.A. 1988, ApJ 333, 24
Cowsik R., McClelland J. 1973, ApJ 180, 7
Edge A.C., Steward G.C., Fabian A.C., Arnaud K.A. 1990, MNRAS 245, 559
Evrard A.E., Davis M. 1988, Nature 333, 335
Evrard A.E. 1989, ApJ 341, L71
Evrard A.E. 1990, ApJ 363, 349
Evrard A.E., Crone M.M 1992, ApJ 394, L1
Frenk C.S., White S.D.M., Davis M. 1983, ApJ 271, 417
Frenk C.S., White S.D.M., Efstathiou G., Davis M. 1990, ApJ 351, 10
Górski K.M., Hinshaw G., Banday A.J., et al. , 1994, Sissa Preprint, astro-ph/9403067
Gunn J.E., Gott J.R. 1972, ApJ 176, 1
Hauser M.G., Peebles P.J.E. 1973, ApJ 185, 757
Henry J.P., Arnaud K.A. 1991, ApJ 372, 410
Holtzman J.A. 1989, ApJS 71, 1
Hugues J.P., Gorenstein P., Fabricant D. 1988, ApJ 329, 82
Hughes J.P. 1992, ApJ 337, 21
Kaiser N. 1983, ApJ 273, L17
Kaiser N. 1986, MNRAS 222, 323
de Lapparent V., Geller M.J., Huchra J.P. 1986, ApJ 302, L1
Lemaître G. 1933, Ann. Soc. Sci. Bruxelles A53, 51
Lilje P.B. 1990, ApJ 351, 1
Melott A.L. 1985, ApJ 289, 2
Melott A.L. 1987, MNRAS 228, 1001
Oukbir J., Blanchard A. 1992, A&A 262, L21
Peacock J.A., Heavens A.F. 1985, MNRAS 217, 805
Peebles P.J.E. 1980, *The large-scale structure of the universe*, Princeton Univ. Press
Peebles P.J.E. 1982, ApJ 258, 415
Press W., Schechter P. 1974, ApJ 187, 425
Smoot G.F., Bennett C.L., Kogut A., *et al.* 1992, ApJ 396, L1
Tuluie R., Matzner R.A., Anninos P. 1994, Sissa Preprint, astro-ph/9403040
White S.D.M., Frenk C.S., Davis M. 1983, ApJ 274, L1
White S.D.M., Davis M., Frenk C.S. 1984, MNRAS 209, 27P
Zeng N., White S.D.M. 1991, ApJ 374, 1This article was processed by the author using Springer-Verlag LaTeX A&A style file *L-AA* version 3.